\ifx\ProceedingsGelomur\relax\else
  \documentclass[11pt,twoside]{book}
  \usepackage[notitlepage]{proceedings}
\def\be{\begin{equation}}
\def\beq{\begin{equation}}
\def\eeq{\end{equation}}
\def\ee{\end{equation}}
\def\bea{\begin{eqnarray}}
\def\eea{\end{eqnarray}}
\def\ba{\begin{array}}
\def\ea{\end{array}}

\def \um {\frac{1}{2}}

\def \ga {\gamma}
\def\rref#1{(\ref{#1})}

\def\ep{\epsilon}

\begin{document}
  \begin{article}
\fi

\title[Running title]{Classical and quantum geometry of moduli spaces 
in three-dimensional gravity}%
\author[Running author(s)]{J.\ E.~Nelson$^1$ and R.\ F.~Picken$^2$}%
\address{$^1$Dipartimento di Fisica Teorica, Universit\`a degli Studi
       di Torino
and Istituto Nazionale di Fisica Nucleare, Sezione di Torino,
via Pietro Giuria 1, 10125 Torino,
Italy\\
 $^2$Departamento de Matem\'{a}tica and CAMGSD -
 Centro de An\'{a}lise Matem\'{a}tica, Geometria e Sistemas Din\^{a}micos,
 Instituto Superior T\'{e}cnico,
Avenida Rovisco Pais, 1049-001 Lisboa, Portugal}%
\maketitle

\begin{abstract} We describe some results concerning the phase space of
$3$-dimensional Einstein gravity when space is a torus  and with negative
cosmological constant. The approach uses the holonomy matrices of flat
$SL(2,\mathbb{R})$ connections on the torus to parametrise the geometry. After
quantization, these matrices acquire non-commuting entries, in such a way that
they satisfy $q$-commutation relations and exhibit interesting geometrical
properties. In particular they lead to a quantization of the Goldman bracket.  
\end{abstract}


\mscMM{83 C 45}

\section{Introduction}

From the point of view of geometry, the theory of classical general relativity
(see Fernando Barbero's lectures in this volume) is the study of Riemannian or
semi-Riemannian geometries (depending on the choice of Euclidean or
Lorentzian signature) which satisfy the Einstein equations. In
$3$-dimensional spacetime these equations for the components $g_{\mu\nu}$ of 
the metric tensor are derived from the Einstein--Hilbert action
\be
\int \sqrt{|g|}(R+\Lambda) d^3x
\label{act}
\ee
where integration is over the spacetime manifold, and we have
included a cosmological constant $\Lambda$. In the first term of \rref{act} 
the Ricci scalar, a contraction of the Riemann tensor, appears. This term may 
be written as follows:
\be
\um \int R^{\mu\sigma}{}_{\rho\nu}\epsilon_{\mu\sigma\alpha}\epsilon^{\rho\nu\beta} 
\delta_{\beta}^{\alpha} d^3x
\label{rie}
\ee
where the usual summation convention over repeated indices is used and indices 
on the totally antisymmetric tensor
$\epsilon_{\mu\nu\rho}$ are raised with the inverse metric tensor
$g^{\mu\nu}$. 

It is convenient to rewrite the action \rref{act} in terms of orthonormal
 dreibeins or triads $e^a$.  These are a local basis of $1$-forms 

\be
e^a = e^a_\mu(x)~ dx^\mu, a=1,2,3
\label{drei}
\ee
such that
\begin{equation}
g_{\mu\nu}dx^\mu \otimes dx^\nu = e^a\otimes e^b \eta_{ab}
\label{dreibein}
\end{equation}
where $\eta_{ab}= {\rm diag}(-1,1,1)_{ab}$. Then the action \rref{act} takes 
the form
\begin{equation}
\int(R^{ab}\wedge e^c + \Lambda e^a\wedge e^b\wedge e^c) \epsilon_{abc}
\label{beinaction}
\end{equation}
where $R^{ab}$ are the curvature $2$-forms 
\be 
R^{ab} = \um R^{ab}{}_{\mu \nu} dx^\mu \wedge dx^\nu
\ee
and $ R^{ab}{}_{\mu \nu}$ is the Riemann tensor that appears in \rref{rie}
contracted with the dreibein components \rref{drei}.

In the dreibein formulation, there is an extra gauge symmetry of local Lorentz
transformations $e^a\mapsto {M^a}{}_b e^b$ where $M\in SO(2,1)$ (local,
since $M$ depends on the point of spacetime). This extra freedom
arises since one may simultaneously rotate the three fields $e^a$,
whilst preserving the metric and the condition \rref{dreibein}.

There is a striking similarity between
the action in the form (\ref{beinaction}) and the Chern-Simons action
for a connection $A$ in a principal $G$-bundle, which has the structure 
\be
\int
(F\wedge A + A \wedge A \wedge A).\nonumber
\ee 
Indeed, it was shown by Witten
\cite{wit} that the action (\ref{beinaction}) may be interpreted as a
Chern-Simons action for $G=SO(2,2)$, when $\Lambda <0$ (and for
$G=SO(3,1)$ when $\Lambda >0$). The connection in the Chern-Simons
theory is given in terms of the dreibein $e^a$ and spin connection (or Ricci
rotation coefficient) $\omega^{ab}$ by:
\be
A=\frac{1}{2}\omega^{ab}M_{ab} + e^aM_{a4}, 
\label{acpt}
\ee
where the indices $a,b$ run from $1$ to $3$, and
$\left\{M_{AB}\right\} _{A,B=1,\dots,4}$ is a basis of the Lie algebra
of $SO(2,2)$. Note that in this so-called first-order formalism, the
dreibein $e^a$ and spin connection $\omega^{ab}$ are independent fields. 

We conclude this introduction with a short discussion of the relation between
connections and holonomy. Given a connection on a principal
$G$-bundle, a holonomy is an assignment of an element $H(\gamma)$ of $G$
to each (based) loop $\gamma$ on the manifold, obtained by lifting the
loop into the total space of the bundle and comparing the starting and
end points of the lifted loop in the fibre over the
basepoint. Holonomy is, in a suitable sense, equivalent to the
connection it is derived from. When the connection is flat, i.e. has
zero fieldstrength $F$, the holonomy of $\gamma$ only depends
on $\gamma$ up to homotopy. Thus an efficient way of describing flat
connections is to specify a group morphism from the fundamental group
of the manifold to the group $G$. 


\section{Equations of motion and the classical phase space\label{sec2}}

Consider the Chern-Simons action 
\be
{\rm tr} \int_{\Sigma\times \mathbb{R}} A\wedge dA + 
\frac{2}{3}A \wedge A \wedge A
\label{cs}
\ee
on a spacetime of the form
$\Sigma\times \mathbb{R}$, where $\Sigma$ is a closed surface
representing space and $\mathbb{R}$ represents time. The connection $1$-form 
$A$ may be written as
\[
A=A_i dx^i + A_0 dx^0
\]
where $x^i,\, i=1,2$ are coordinates on $\Sigma$, and $x^0=t$ is the
time coordinate. Imposing the gauge fixing condition
\[
A_0=0
\]
and the corresponding constraint
\[
F_{ij}dx^i \wedge dx^j =0 
\]
we see that the connections are flat. The action \rref{cs} now has the structure
\[
\int A_2 \partial_0 A_1 + \dots 
\]
and therefore ``$A$ is its own conjugate momentum''. The Poisson
brackets for the components of $A$ (see equation \rref{acpt}) have the 
following form:
\begin{equation}
\left\{A_1{}^a(x), A_j{}^b(y) \right\} = \delta^{ab}\ep_{ij}\delta^2 (x - y)
\quad \ep_{12}=1.
\label{pb}
\end{equation}

We now choose the space manifold to be the torus $\mathbb{T}^2$, and since the
group $SO(2,2)$ is isomorphic, up to a discrete identification, to the product
$SL(2,\mathbb{R})\times SL(2,\mathbb{R})$, we restrict ourselves to
studying the phase (moduli) space of flat $SL(2,\mathbb{R})$ connections on the
torus $\mathbb{T}^2$, modulo gauge transformations. Note that this is in
principle a complicated space to describe, being an infinite-dimensional space
divided by an infinite-dimensional group, but in the holonomy picture there is a
very simple finite-dimensional description. 

Since $\pi_1(\mathbb{T}^2)=\left< \gamma_1, \gamma_2 |
  \gamma_1 \gamma_2 \gamma_1^{-1} \gamma_2^{-1}=1\right>$ 
where $\gamma_1$ and $\gamma_2$ are a pair of generating cycles, a holonomy 
\[
H:\pi_1(\mathbb{T}^2)\rightarrow SL(2,\mathbb{R})
\] 
is given by $U_1:=H(\gamma_1)$ and $U_2:=H(\gamma_2)$, since this
determines $H$ on any other homotopy class of loops. The phase space
$P$ is then 
\[
P=\left\{(U_1,U_2)|U_1U_2=U_2U_1\right\}/\sim
\]
where $\sim$ denotes the remaining gauge freedom, namely 
\[
(U_1,U_2)\sim (S^{-1}U_1S, S^{-1}U_2S)
\]
for any $S\in SL(2,\mathbb{R})$. 

For a single matrix $U\in SL(2,\mathbb{R})$ there are four
possibilities for how $U$ can be conjugated into a standard form:
\begin{itemize}
\item[A)] $U$ has $2$ real eigenvalues:
\[
S^{-1}US=
\left(\begin{array}{ll}\lambda & 0 \\ 0 & \lambda^{-1}\end{array}\right)
\]
\item[B)]  $U$ has $1$ real eigenvalue with an eigenspace of
  dimension $2$:
\[
S^{-1}US= \pm \left(\begin{array}{ll} 1& 0 \\0&1
\end{array}\right)
\]
\item[C)] $U$ has $1$ real eigenvalue with an eigenspace of
  dimension $1$:
\[
S^{-1}US=
\left(\begin{array}{ll}\pm 1 & 1 \\ 0 & \pm 1\end{array}\right)
\]
\item[D)] $U$ has no real eigenvalues:
\[
S^{-1}US=
\left(\begin{array}{ll}\cos \theta & -\sin \theta \\ \sin \theta & \cos
\theta \end{array}\right)
\]
\end{itemize}

A similar analysis for a pair of commuting
$SL(2,\mathbb{R})$ matrices led in \cite{mod} to an explicit
parametrization of the classical phase space $P$. Its structure
resembles that of a cell complex with, for instance $2$-dimensional cells
consisting of pairs of diagonal matrices, or pairs of rotation
matrices. However there are also $1$-dimensional cells which consist of
e.g. pairs of non-diagonalisable matrices of the form:
\begin{equation}
U_1= 
\left( \begin{array}{cc} 1 &\cos \alpha\\ 0&1\end{array}
\right) \quad
U_2= 
\left( \begin{array}{cc} 1 &\sin \alpha\\ 0 & 1
\end{array} \right), \, 0< \alpha < \frac{\pi}{2}.
\label{CC}
\end{equation}
For further details and depictions of $P$ see \cite{mod}.

\section{Quantization via quantum matrices\label{sec3}}

The Poisson brackets (\ref{pb}) are for non-gauge-invariant variables
so it is convenient to change to gauge-invariant variables, and an
obvious choice are the traced holonomies
\[
T(\gamma) = \um ~{\rm tr}~ H(\gamma)
\]
which are gauge-invariant due to the conjugation invariance of the
trace. The holonomy is sometimes written as a path-ordered
exponential, or Chen integral,
\[
H(\gamma)= {\cal P} \exp \int_\gamma A
\]
and from equation \rref{pb} the Poisson brackets
between the $T(\gamma)$ are only non-vanishing if the loops
intersect transversally. From trace identities for $2 \times 2$
matrices it is enough to consider the following three variables: 
\[
T_1:=T(\gamma_1) \quad T_2:=T(\gamma_2) \quad T_3:=T(\gamma_1\gamma_2) 
\]
(which are not independent since they satisfy the identity 
$T_1^2 + T_2^2 + T_3^2 -2T_1T_2T_3=1$). Their Poisson bracket relations 
are \cite{nr}
\begin{equation}
\left\{T_i, T_j\right\} = \ep_{ij}{}^k T_k + T_i T_j, \quad i,j,k=1,2,3.
\label{pbT}
\end{equation}
(here we have rescaled the variables compared to \cite{nr} to absorb the coupling
constants). 

The first term on the right-hand-side of equation \rref{pbT} has a geometric
interpretation in terms of rerouted loops: e.g. for $i=1,\,j=2$ the two cycles
$\gamma_1$ and $\gamma_2$ intersect transversally at one point, and from
homotopy invariance of the holonomy $T_3$ is the traced holonomy corresponding
to the loop $\gamma_1 S \gamma_2$ obtained
by starting at the basepoint, following $\gamma_1$ to the intersection point 
$S$,
rerouting along the loop $\gamma_2$ back to the intersection point,
and finally continuing again along $\gamma_1$ back to the basepoint. We will
see more of these rerouted loops shortly.

We observe that by parametrising the variables as follows:
\[
T_1 = \cosh r_1 \quad T_2 = \cosh r_2 \quad T_3= \cosh (r_1+r_2)
\]
equation (\ref{pbT}) is solved by setting:
\[
\left\{ r_1, r_2\right\} = 1.
\]
On quantization, replacing $T_i,\,r_j$ by operators 
$\hat{T}_i,\, \hat{r}_j$ respectively implies the corresponding commutation 
relation:
\begin{equation}
[\hat{r}_1, \hat{r}_2] = i \hbar.
\label{rcommrel}
\end{equation}

The operators $\hat{T}_i$ satisfy a $q$-deformed ($q=e^{i\hbar}$) cubic 
relation, which can be interpreted in terms of a quantum Casimir operator for 
the quantum group $SU(2)_q$ - see \cite{nrz}.

We note that e.g.
\[
T_1=\um ~{\rm tr}~ U_1=\cosh r_1 =\um (e^{r_1} + e^{-r_1})
\]
so that by introducing the {\it quantum} matrices
\begin{equation}
\hat{U}_i=
\left(\begin{array}{ll} e^{\hat{r}_i} & 0 \\ 0 & e^{-\hat{r}_i}\end{array}\right)
\, i=1,2
\label{Ui}
\end{equation}
we have the analogous relation between $\hat{T}_i$ and $\hat{U}_i$, namely
\[
\hat{T}_i = \um ~{\rm tr}~ \hat{U}_i \quad i=1,2.
\]
We also notice that these quantum matrices satisfy the following fundamental
relation:
\begin{equation}
\hat{U}_1\hat{U}_2 = q~\hat{U}_2 \hat{U}_1,
\label{fund}
\end{equation}
where we are using matrix multiplication of operator-valued matrices (the usual 
algebraic rule, but paying strict attention to the order of the symbols). For
example, the relation 
\[
e^{\hat{r}_1} e^{\hat{r}_2} = q~ e^{\hat{r}_2}e^{\hat{r}_1}
\]
follows from the commutation relation 
\rref{rcommrel} between the operators $\hat{r}_i$.   

The cubic constraint satisfied by the quantum variables $\hat{T}_i$ is rather 
complicated, so instead we work with the quantum holonomy matrices 
$\hat{U}_i$ themselves rather than with the trace functions $\hat{T}_i$. It is
important to note that even though the quantum matrices $\hat{U}_i$ are 
not gauge-invariant, i.e.
\[
\hat{U}_i \neq S^{-1}\hat{U}_i S
\]
for general $S$, the fundamental equation (\ref{fund}) is gauge-covariant, and
is also covariant under the modular symmetry of the theory, i.e. the group of 
large diffeomorphisms of the torus - see \cite{np1}. Thus
our idea is to substitute invariant variables obeying complicated equations by
non-invariant matrix variables satisfying natural  $q$-commutation relations
like the fundamental relation (\ref{fund}). Certainly for the case of diagonal
matrices these two viewpoints are entirely equivalent.

We have also studied, in \cite{np1}, what happens when one imposes the
fundamental equation for a pair of upper-triangular quantum matrices, which
should correspond, in some sense, to the quantization of the $1$-dimensional
upper-triangular cell of the classical phase space mentioned in section
\ref{sec2}. 

If one parametrizes the quantum matrices $\hat{U}_i$ as follows:
\begin{equation}
\hat{U}_i = \left(\begin{array}{ll} \hat{\alpha}_i & \hat{\beta}_i 
\\ 0 & \hat{\alpha}_i^{-1}\end{array}\right),
\label{upptriang}
\end{equation}
where the $\hat{\alpha}_i, \, \hat{\beta}_i$ are operators to be determined, a
solution to equation \rref{fund} is given by:
\begin{eqnarray}
\hat{\alpha}_1 \psi(b) & = & \exp \frac{d}{db} \psi(b) \nonumber \\
\hat{\alpha}_2 \psi(b) & = & \exp i\hbar b\, \psi(b) \nonumber \\
\hat{\beta}_i \psi(b) & = & \hat{\alpha}_i \psi(-b)
\label{sol}
\end{eqnarray}
Note the change of sign in the argument of $\psi$ in the last of equations
\rref{sol}. It can be checked, from \rref{sol} that  
\[
\hat{\alpha}_1 \hat{\alpha}_2 = q ~\hat{\alpha}_2 \hat{\alpha}_1
\]
as required, but we also get an {\it internal} commutation relation
\[
\hat{\alpha}_1 \hat{\beta}_1 =  \hat{\beta_1} \hat{\alpha}_1^{-1}
\]
for the elements of $\hat{U}_1$ and similarly for $\hat{U}_2$, which curiously 
does not involve the quantum parameter $\hbar$. 

Note that it is impossible to  find solutions to \rref{fund} with
$\hat{\alpha}_i = \mathbb{I}$, the unit operator, by naive analogy with
equation \rref{CC}, since $\hat{\beta}_1 + \hat{\beta}_2 \neq q (\hat{\beta}_1
+ \hat{\beta}_2)$. Thus in terms of the number of quantum parameters, this
upper-triangular sector would appear to be  as substantial as the triangular
sector, unlike the classical case. 

Finally we remark that in \cite{qmp}, we studied equations like
(\ref{upptriang}) from an algebraic point of view, and found that their
solutions have several interesting properties analogous to quantum groups. 


\section{Reroutings and the quantized Goldman bracket}

Here we briefly describe our most recent work - for a full treatment see
\cite{goldman}. In section \ref{sec3} we only considered the quantum matrices
assigned to $\gamma_1$ and $\gamma_2$, so it is natural to try and understand
how to assign quantum matrices to other loops, and to study the relationships 
between them. A useful way of doing this, proposed in 
\cite{mikpic}, is to introduce a constant quantum
connection
\[
\hat{A}= (\hat{r}_1 dx + \hat{r}_2 dy) 
\left(\begin{array}{ll} 1& 0 \\0& -1 \end{array}\right),
\]
where constant means that the $\hat{r}_i$ do not depend on the spatial 
coordinates $x,\,y$ of the torus. Then the assignment of a quantum matrix to 
any loop is given by the holonomy of this connection along the loop:
\be
\gamma \mapsto \hat{U}_\gamma = \exp \int_{\gamma} \hat{A}.
\label{hol}
\ee
It can easily be seen that \rref{hol} reproduces the quantum matrices
$\hat{U}_i$ of equation (\ref{Ui}), if $\gamma_1$ is the loop with $y$
coordinate constant and $x$ running from $0$ to $1$, and $\gamma_2$ is the loop
with $x$ constant  and $y$ running from $0$ to $1$. 

It is convenient to identify the torus $\mathbb{T}^2$ with $\mathbb{R}^2 /
\mathbb{Z}^2$, where $\mathbb{Z}^2$ consists of points with integer $x$ and $y$
coordinates. We consider all loops on the torus represented by
piecewise-linear (PL) paths between integer points on the $(x,y)$ plane, and
work with this description, keeping in mind that paths represent loops. In
particular the integer paths denoted $(m,n)$ are straight paths
between $(0,0)$ and $(m,n)$ with $m,\,n$ integers. Thus for example we assign
to the integer path $(2,1)$ the quantum matrix  
\[
\hat{U}_{(2,1)}= 
\left(\begin{array}{ll} e^{2\hat{r}_1+ \hat{r}_2} & 0 
\\ 0 & e^{-2\hat{r}_1 -\hat{r}_2}
\end{array}\right).
\]

Consider two homotopic loops $\gamma_1$ and $\gamma_2$ corresponding to
PL paths both starting at $(0,0)$ and ending at the same integer point in the
plane. It was shown in \cite{goldman} that there is the following relationship
between the respective quantum matrices:
\begin{equation}
\hat{U}_{\gamma_1}=q^{S(\gamma_1,\gamma_2)}\hat{U}_{\gamma_2},
\label{areaphase}
\end{equation}
where $S(\gamma_1,\gamma_2)$ denotes the signed area enclosed between the paths
$\gamma_1$ and $\gamma_2$. For example, the exponent (the number $1$) of $q$ 
in the fundamental relation \rref{fund} is the signed area between two paths
around the perimeter of the unit square, starting at $(0,0)$ and ending at
$(1,1)$, the first via $(1,0)$ and the second via $(0,1)$. 

The traces of these quantum matrices also exhibit commutation relations with
interesting properties. Let 
\[
\hat{T}(m,n) := {\rm tr} ~\hat{U}_{(m,n)}. 
\]
(note we have dropped the factor $\frac{1}{2}$ for easier comparison with the
Goldman result below). 
It was shown in \cite{goldman} that the following commutation relation holds:
\begin{equation}
[\hat{T}(m,n), \hat{T}(s,t)]=
(q^{\frac{mt-ns}{2}}-q^{-\frac{mt-ns}{2}}) \left(\hat{T}(m+s,n+t) - \hat{T}(m-s,n-t)\right)
\label{qgb1}
\end{equation}

There are some surprising geometric aspects to equation \rref{qgb1}. The number
$mt-ns$ appearing in the exponents is the signed area of the parallelogram
spanned by the vectors $(m,n)$ and $(s,t)$. The same expression equals the
suitably-defined total intersection number (including and counting
multiplicities) of the two loops represented by the paths $(m,n)$ and $(s,t)$.
Equation \rref{qgb1} can, in fact, be viewed as a quantization of a bracket due
to Goldman \cite{wmgoldman} for the loops corresponding to such integer paths.
This bracket is a Poisson bracket for the functions 
$T(\gamma)={\rm tr}\, U_\gamma$ given by:
\be 
\{T(\ga_1), T(\ga_2)\} = \sum_{S \in \ga_1 \sharp \ga_2}
\epsilon(\ga_1,\ga_2,S)(T(\ga_1S\ga_2) - T(\ga_1S\ga_2^{-1})) 
\label{gold} 
\ee
where $\gamma_1 \sharp \gamma_2$ denotes the set of transversal intersection
points of $\gamma_1$ and $\gamma_2$ and $\epsilon(\gamma_1,\gamma_2,S)$ is
their intersection index for the intersection point $S$. In equation
\rref{gold} $\ga_1S\ga_2$ and $\gamma_1S\gamma_2^{-1}$ denote 
the loops which follow $\gamma_1$
and are rerouted along $\gamma_2$, or its inverse, at the intersection point
$S$  as described previously. For
the integer loops considered here, all the rerouted loops $\gamma_1S\gamma_2$
are homotopic to the integer loop $(m+s, n+t)$, with an analogous statement for
the loops $\gamma_1S\gamma_2^{-1}$. It follows that the classical Goldman
bracket \rref{gold} can be written as
\[ \left\{T(m,n), T(s,t)\right\} = (mt-ns)
(T(m+s,n+t)- T(m-s,n-t)). \] 
Therefore the first factor on the right hand side of (\ref{qgb1}) may be
thought of as a quantum total intersection number for the loops $(m,n)$ and
$(s,t)$.

We remark that in \cite{goldman} we also derived a different form of
\rref{qgb1} where each rerouted loop appears separately. The different terms
are related by the same area phases as in (\ref{areaphase}). In these proofs we
used a classical geometric result \cite{pick} namely Pick's
Theorem (1899), which expresses the area $A(P)$ of a lattice polygon $P$ with
vertices at integer lattice points of the plane in terms of the number of
interior lattice points $I(P)$ and the number of boundary lattice points $B(P)$
as follows:
\[
A(P) = I(P) + \frac{B(P)}{2} -1.
\]
Full details are given in \cite{goldman}.


\section*{Acknowledgments}

 This work was supported by the Istituto Nazionale di Fisica Nucleare (INFN) of
Italy, Iniziativa Specifica FI41, the Italian Ministero dell'Universit\`a e
della Ricerca Scientifica e Tecnologica (MIUR), and by the programme {\em
Programa Operacional ``Ci\^{e}ncia, Tecnologia, Inova\c{c}\~{a}o''} (POCTI) of
the  {\em Funda\c{c}\~{a}o para a Ci\^{e}ncia e a Tecnologia} (FCT), 
cofinanced by the European Community fund FEDER.

\ifx\ProceedingsGelomur\relax
  \endinput
\else
\end{article}
\end{document}
\fi